\documentclass[pra,showpacs,twocolumn]{revtex4}

\newcommand{\ket}[1]{|{#1}\rangle}
\newcommand{\bra}[1]{\langle{#1}|}

\newcommand{\be}{\begin{equation}}
\newcommand{\ee}{\end{equation}}

\usepackage{amsfonts}
\usepackage{amsmath}
\usepackage{graphicx}
\usepackage{amssymb}
\usepackage{amsmath}
\usepackage{amssymb}
\usepackage{graphicx}
\usepackage{lscape}
\usepackage{color}
\usepackage{epstopdf}

\begin{document}
\title{Three-body repulsive forces among identical bosons in one dimension}

\author{M. Valiente}
\affiliation{Institute for Advanced Study, Tsinghua University, Beijing 100084, China}

\begin{abstract}
I consider non-relativistic bosons interacting via pairwise potentials with infinite scattering length and supporting no two-body bound states. To lowest order in effective field theory, these conditions lead to non-interacting bosons, since the coupling constant of the Lieb-Liniger model vanishes identically in this limit. Since any realistic pairwise interaction is not a mere delta function, the non-interacting picture is an idealisation indicating that the effect of interactions is weaker than in the case of off-resonant potentials. I show that the leading order correction to the ground state energy for more than two bosons is accurately described by the lowest order three-body force in effective field theory that arises due to the off-shell structure of the two-body interaction. For natural two-body interactions with a short-distance repulsive core and an attractive tail, the emergent three-body interaction is repulsive and, therefore, three bosons do not form any bound states. This situation is analogous to the two-dimensional repulsive Bose gas, when treated using the lowest-order contact interaction, where the scattering amplitude exhibits an unphysical Landau pole. The avoidance of this state in the three-boson problem proceeds in a way that parallels the two-dimensional case. These results pave the way for the experimental realisation of one-dimensional Bose gases with pure three-body interactions using ultracold atoms. 
\end{abstract}
\pacs{
}
\maketitle
\section{Introduction}
The theory of few-particle forces in quantum mechanics has a long history that dates back to the early studies of atomic nuclei \cite{Yukawa}. It was soon realised that even highly sophisticated nucleon-nucleon potentials, which faithfully reproduced all experimental features of the deuteron \cite{Erkelenz} and the nucleon-nucleon scattering amplitudes \cite{Bryan}, failed to account for the binding energy of the triton \cite{Bomelburg}. Tuning the short distance details of the nuclear potential, affecting the off-shell elements of the two-nucleon amplitude, moreover, is unnecessary, since these are not measurable and can be traded off in favour of on-shell three-body amplitudes \cite{ReviewThreeBody}. This is where three-body forces come into play, as they can be used, in conjunction with accurate two-body interactions, to fit three-nucleon data \cite{Nogga}, so that heavier nuclei can be investigated in this way. 

The modern theory of few-body forces has evolved into a systematic, well controlled low-energy expansion of the interparticle interactions \cite{Epelbaum1,Epelbaum2}. Based on the pioneering work of Weinberg on effective nuclear forces \cite{WeinbergEFT1,WeinbergEFT2}, model independent two- and higher-body interactions have been developed into what is now commonly known as (chiral) effective field theory (EFT). In essence, EFT considers all possible interactions that are consistent with the underlying symmetries of the problem at a given order in perturbation theory, and the bare coupling constants of the theory are renormalised in favour of low-energy, physical observables. 

These effective theories, which are commonplace in nuclear physics, and to a lesser extent in the physics of cold Helium \cite{Kievsky}, have slowly made their way into the ultracold atomic realm \cite{BraatenEFT}. In fact, the lowest-order interactions were first introduced in the theory of Bose-Einstein condensates (BECs) using pseudopotentials back in 1957 \cite{HuangYang}. Since the original motivation in ultracold gases was to produce BECs with alkali atoms \cite{BEC1,BEC2}, which interact very weakly, higher-order EFTs were unnecessary for a long time. Three-body interactions in three spatial dimensions, however, were shown to be needed in order to fix the energy of the lowest-lying Efimov state at or near unitarity and avoid the Thomas collapse \cite{Bedaque1,Bedaque2}, thereby generating great interest in few-body forces in the atomic physics community, which saw the first experimental evidence \cite{Grimm,Grimm2,Zaccanti} for the elusive Efimov states \cite{Efimov}. Within this context, repulsive three-body forces have been proposed as a mechanism for the stabilisation of quantum atomic droplets \cite{Bulgac} which, however, turned out to be stabilised by quantum fluctuations -- a lowest-order effect in EFT -- at least in actual experimental demonstrations \cite{Tarruell,Inguscio}. The emergence of effective multiparticle forces in an ultracold atomic setting have been studied mostly in three dimensions. Up to five-body forces emerging from two-body EFT including the scattering length and effective range in external trapping potentials have been calculated in three dimensions \cite{Yin,Johnson1,Johnson2}, where a non-universal (i.e. beyond two-body EFT) three-body force was also shown to be necessary to regularise and renormalise the problem. Interestingly, effects of multiparticle interactions, stemming from a single well of an optical lattice \cite{Johnson1}, have been observed experimentally \cite{Will}. Other instances of multiparticle effective terms, not related to asymptotic low-energy physics,  include three-body correlated tunneling in double wells \cite{Dobr} or many-body coupling in trapped ionic systems \cite{Bermudez}. 

The most promising candidate for the observation of effects due to pure three-particle forces, disentangled from other typically more relevant two-body effects, is perhaps a system of ultracold bosons tightly confined to one spatial dimension. Recently, Guijarro \textit{et al.} proposed using Bose-Bose and Fermi-Bose mixtures to engineer three-body repulsive interactions between dimers \cite{PetrovThreeBody}. This proposal relies upon the ability to independently and simultaneously tune two different intraspecies interaction strengths, besides the interspecies scattering length. There are also several recent works focusing on attractive three-body forces in one dimension without \cite{Drut1,Drut2,Nishida2} and with \cite{Nishida3BodyLattice} a reference to a physical implementation, the latter requiring simultaneous tuning of several interaction strengths in a multicomponent Bose system on a tight-binding optical lattice. The trimer may also be observable with trapped ultracold atoms, as shown by Pricoupenko \cite{Pricoupenko1}, who also developed the pseudopotential treatment of the three-body interaction in Ref.~\cite{Pricoupenko}, which is most convenient for studies in the position representation. In the thermodynamic limit, mean-field and beyond-mean-field corrections to the ground state energy of the one-dimensional Bose gas with pure three-body repulsive interactions were recently obtained by Pastukhov \cite{PastukhovThreeBody}. Interestingly, strong three-body interactions, if these can be engineered in one dimension, may give rise to rather exotic (anyonic) exchange statistics that are governed by the traid group \cite{Harshman}, instead of the traditional braid group \cite{anyons}.

What all of the above works on the one-dimensional three-body interaction agree upon is the important fact that the three-body problem with pure three-body forces in one dimension is kinematically equivalent to a two-dimensional two-body problem at low energies. Indeed, the former exhibits the same quantum anomaly as the latter, which has recently been investigated experimentally in two different works \cite{Selim, Vale}, a fact that was the focus of Refs.~\cite{Drut1,Drut2} in the present case. The most immediate consequence of this is that, while for attractive interactions three- and many-body bound states appear, for repulsive interactions one needs to deal with an unphysical bound state, i.e. a Landau pole in the scattering amplitude. Fortunately, it is possible to deal with it in the same way as for the two-dimensional problem with two-body interactions thanks to the kinematic equivalence. 

\section{Two-body interactions}
I consider non-relativistic identical bosons interacting via two-body forces exhibiting a zero-energy resonance (infinite scattering length \footnote{In the literature, there is no general consensus in the terminology for infinite scattering length. This situation is sometimes referred to as "zero crossing". I shall, however, call this a zero-energy resonance, in analogy with the three-dimensional case.}).
The model interactions I use have a soft repulsive core at short distances, and an attractive finite-range tail. This type of interactions are justified for effectively reduced-dimensional systems after integrating out the transversal degrees of freedom \cite{AdrianDelMaestro1,AdrianDelMaestro2,AdrianDelMaestro3}. I shall use two different forms of the two-body interaction. The first interaction potential $V(x)=V(x_i-x_j)$ between particles $i$ and $j$ that I will use is given by
\begin{equation}
  V(x)=V_0e^{-\lambda_0x^2} + V_1e^{-\lambda_1x^2},\label{potential}
\end{equation}
where $V_0$ ($<0$) and $V_1$ ($>0$) give, respectively, the strength of the attractive tail and soft core of the interaction, $\lambda_0$ and $\lambda_1$ determine their spatial spread, and I shall denote by $x_0$ ($x_0^2=\log|\lambda_1V_1/\lambda_0V_0|/(\lambda_1-\lambda_0)$) the length scale determining the potential minimum. The second type is given by $V(x)=W(x)+W(-x)$, with
\begin{equation}
  W(x)=\frac{V_0}{2\cosh(\sqrt{\lambda_0}x)}+\frac{V_1}{2\cosh(\sqrt{\lambda_1}x-b)}.\label{COSH}
\end{equation}
The minimum of this potential will be also denoted by $x_0$. In what follows, I choose these parameters in such a way that the two-boson scattering length diverges ($1/a=0$), i.e., such that the zero-energy solution to the stationary two-body Schr{\"o}dinger equation in the relative coordinate
\begin{equation}
  -\frac{\hbar^2}{m}\psi"(x) + V(x) \psi(x) = 0,
\end{equation}
  has the asymptotic form $\psi(x)\propto 1$ as $x\to \pm \infty$, and such that there are no two-body bound states. The effective two-body interaction, to lowest order and in the momentum representation, is given by a vanishing Lieb-Liniger coupling constant $g=-2\hbar^2/ma = 0$ \cite{LiebLiniger}.

  In the two-boson sector, the next-order interaction involves the effective range $r$ \cite{ValienteZinnerEFT}, whose effect is identically zero at zero energy. To see this, and to analyse the three-body problem, it is most convenient to abandon the collision-theoretical approach and instead place the few-body systems on a finite line of length $L$ with periodic boundary conditions. The analysis of the finite-size spectrum can be used to extract low-energy scattering amplitudes \cite{Luscher1,Luscher2}, and has come to be the method of choice in modern studies of scattering processes, from low-energy nuclear physics \cite{BeaneTwoNucleons}  to lattice QCD \cite{LatticeQCDReview}. In 1D, the eigenenergies $E=\hbar^2k^2/m$ at zero total momentum for two-bosons can be calculated from the equation
  \begin{equation}
    k=\frac{2\pi n}{L} - \frac{2}{L}\theta(k),\hspace{0.2cm} n\in\mathbb{Z},
  \end{equation}
where $\theta(k)$ is the even-wave scattering phase shift in 1D \cite{Adhikari1D}. For the ground state ($n=0$), since $1/a=0$, we obtain the solution $k=0$ and therefore, as claimed, the effective range has no effect on it. For the first excited state ($n=1$), however, using $k\tan\theta(k) = 1/a + rk^2/2 + O(k^4)$ \cite{Adhikari1D}, the energy shift with respect to the non-interacting energy $E_{1}^{(0)}$ is given by $\Delta E_{1} \approx -2E_1^{(0)} r /L = O(L^{-3})$. Therefore, the lowest-order correction for $N\ge 3$ particles is given by the contribution of effective three-body forces which na{\"i}vely scales as $O(L^{-2})$, for both ground and low-lying excited states. These come from off-shell components of the two-body transition matrix, as shown in Appendix \ref{apendicitis}.

\section{Effective three-body forces}
  The bare lowest-order three-body interaction $V_3^{\mathrm{LO}}$ is obtained by expanding a 1D hyperspherically symmetric three-body potential to zero-th order in the hyperspherical momentum, and corresponds to a contact interaction in the position representation of the form
  \begin{equation}
    V_3^{\mathrm{LO}}(x_1,x_2,x_3) = g_3 \delta(x_1-x_2)\delta(x_2-x_3),\label{V3LO}
  \end{equation}
  where $g_3$ is the bare three-body interaction strength. For a pure three-body interaction, the three-body scattering amplitude can be obtained directly through the Lippmann-Schwinger equation since the Faddeev decomposition is unnecessary. The three-body T-matrix $\hat{T}_3(z)$, with $z$ the (complex) collision energy, for the interaction (\ref{V3LO}) is readily obtained by solving the Lippmann-Schwinger equation
\begin{equation}
  \hat{T}_3(z)=\hat{V}_3^{\mathrm{LO}}+\hat{V}_3^{\mathrm{LO}}\hat{G}_0(z)\hat{T}_3(z)
\end{equation}
where $\hat{G}_0(z)=(z-\hat{H}_0)^{-1}$ is the non-interacting Green's function, as $\bra{k_1',k_2',k_3'}\hat{T}_3(z)\ket{k_1,k_2,k_3} = 2\pi \delta(K-K') t_3(z)$, with $K=k_1+k_2+k_3$ and $K'=k_1'+k_2'+k_3'$ the conserved total momentum. The constant $t_3(z)$, after setting the total momentum to zero, is given by
  \begin{equation}
    t_3(z) =\left[g_3^{-1} - \mathcal{I}(z)\right]^{-1},\label{t3z}
  \end{equation}
  where
  \begin{equation}
\mathcal{I}(z)=
\int \frac{\mathrm{d}q_1\mathrm{d}q_2\mathrm{d}q_3}{(2\pi)^2} \frac{\delta(q_1+q_2+q_3)}{z-\frac{\hbar^2}{2m}(q_1^2+q_2^2+q_3^2)}.\label{Iz}
  \end{equation}
  In order to calculate the coupling constant $g_3$, the integral $\mathcal{I}(z)$, Eq.~(\ref{Iz}), must be regularised. I use a hard cutoff $\Lambda$ in the hyperradial integral, by changing variables to Jacobi coordinates $x=(q_1-q_2)/\sqrt{2}$, $y=\sqrt{2/3}\left[q_3-(q_1+q_2)/2\right]$, and defining the hyperradial momentum $\rho=\sqrt{x^2+y^2}$. The real part of $\mathcal{I}(z)$ for $z=E+i0^{+}$ ($E>0$) is given, in the limit $\Lambda\to \infty$, by
  \begin{equation}
\mathrm{Re}\mathcal{I}(z) = -\frac{m}{2\pi \sqrt{3} \hbar^2} \log\left|\frac{\Lambda^2}{2mE/\hbar^2} \right|\label{ReI}
  \end{equation}
  For attractive interactions, the T-matrix is renormalised by fixing the three-body binding energy $E_B=-|E| \equiv \hbar^2Q_*^2/2m$ while, for repulsion, $E_B$ marks the location of a (unphysical) Landau pole, completely equivalent to its two-body two-dimensional counterpart \cite{Beane2D}. Here, $Q_*$ plays the role of a momentum scale beyond which the EFT description breaks down. As noted by Beane in Ref.~\cite{Beane2D} for the 2D case, the three-body scattering length \cite{PetrovThreeBody} is not a natural scale for repulsive interactions and I shall refer to the momentum scale $Q_*$ only. I continue Eq.~(\ref{Iz}) analytically to negative energies $z<0$ and, setting the location of the pole of $t_3(z)$, Eq.~(\ref{t3z}), to $E_B$, the coupling constant $g_3=g_3(\Lambda)$ is given by
  \begin{equation}
    \frac{1}{g_3(\Lambda)} = \frac{m}{\sqrt{3}\pi \hbar^2} \log\left|\frac{Q_*}{\Lambda}\right|.
  \end{equation}

\section{Finite-size scaling in the three-body problem}  
  Since I will analyse the three-body problem on a two-body resonance using diagonalisation in a periodic box, I derive now the finite-size scaling of the three-body energy with three-body interactions. This is easiest to do in the momentum representation. The stationary Schr{\"o}dinger equation $(\hat{H}_0+\hat{V}_3^{\mathrm{LO}})\ket{\psi} = E \ket{\psi}$, with $\hat{H}_0$ the non-relativistic kinetic energy operator for three particles, is solved by finding the poles of the Green's function for total momentum $K=0$ in a box of length $L$. After writing the energy as $E=(2\pi)^2\hbar^2\lambda^2/mL^2$, and defining an integer cutoff $n_{\Lambda}$ via $\Lambda = 2\pi n_{\Lambda}/L$, the following equation for $\lambda^2$ is found
  \begin{equation}
\sum_{n_1,n_2}'\frac{1}{n_1^2+n_2^2+n_1n_2-\lambda^2}-\frac{4\pi}{\sqrt{3}}\log n_{\Lambda} +\frac{4\pi}{\sqrt{3}} \log \left|\frac{Q_*L}{2\pi}\right| = 0,\label{exactequation}
    \end{equation}
  where the primed sum restricts the values of $(n_1,n_2)$ to $n_1^2+n_2^2+n_1n_2<n_{\Lambda}^2$, and the limit $n_{\Lambda}\to \infty$ is implied. From Eq.~(\ref{exactequation}), it is simple to extract the weak-coupling ($\lambda^2 \ll 1$) expansion, as in previous EFT-based approaches to the finite-size spectrum for two interacting particles \cite{BeaneTwoNucleons,Beane2D,ValienteZinnerL}. Expanding the sum in Eq.~(\ref{exactequation}), I find
  \begin{align}
    &\sum_{n_1,n_2}'\frac{1}{n_1^2+n_2^2+n_1n_2-\lambda^2}-\frac{4\pi}{\sqrt{3}}\log n_{\Lambda}= \nonumber\\
    &-\frac{1}{\lambda^2}+\sigma_1+\sum_{j=1}^{\infty}\sigma_{j+1}\lambda^{2j},\label{sumitas}\\
    \sigma_1&=\sum_{\mathbf{n}\ne \mathbf{0}}'\frac{1}{n_1^2+n_2^2+n_1n_2}-\frac{4\pi}{\sqrt{3}}\log n_{\Lambda},\\
    \sigma_{j}&=\sum_{\mathbf{n}\ne \mathbf{0}}\frac{1}{(n_1^2+n_2^2+n_1n_2)^j}.
  \end{align}
  The values of the first two sums above are calculated to be $\sigma_1=3.96156\ldots$, $\sigma_2=8.7115\ldots$. Using the expansion (\ref{sumitas}) of the sum in Eq.~(\ref{exactequation}) to obtain a weak-coupling expansion in the renormalised coupling constant $g_R$, given by \footnote{As in the two-body problem in two dimensions, there is freedom in choosing the renormalised coupling constant, i.e. by choosing different scales \cite{Beane2D}. This amounts to a mere reparametrisation of the weak-coupling expansion.}
\begin{equation}
  g_R=\frac{\sqrt{3}}{4\pi}\frac{1}{\log\left|\frac{Q_*L}{2\pi}\right|},\label{gR}
\end{equation}
the ground state energy $E_0$ of the three-body system reads
  \begin{equation}
    E_0 = \frac{4\pi^2\hbar^2}{mL^2}\left[g_R-\sigma_1g_R^2+(\sigma_1^2-\sigma_2)g_R^3 + O(g_R^4)\right].\label{E0}
  \end{equation}
As seen above, the na{\"i}ve scaling of the energy ($\propto L^{-2}$) is modified by the quantum anomaly \cite{Drut1,Drut2} in the form of logarithmic corrections. 

\begin{figure}[t]
\includegraphics[width=0.5\textwidth]{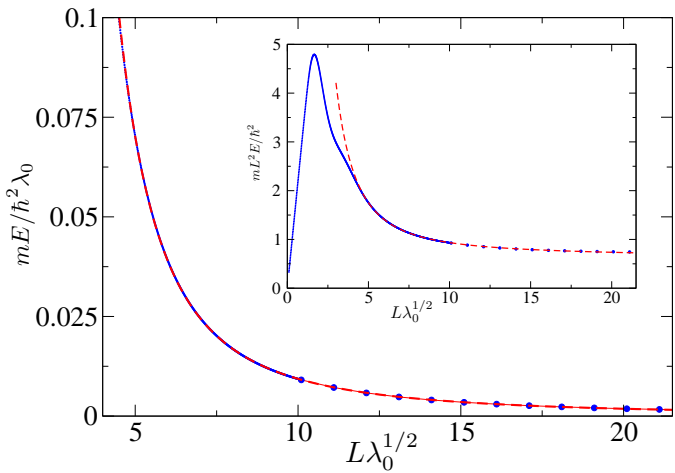}
\caption{Ground state energy of three particles with pairwise interactions in Eq.~(\ref{potential}), with $\lambda_0^{1/2}x_0 = 1.076\ldots$, $\lambda_1/\lambda_0 = 2$, $mV_0/\hbar^2\lambda_0=-5$, $V_1/V_0=-1.59151239$, corresponding to inverse scattering length $x_0/a \approx -3.5\cdot 10^{-7}$. Small blue dots correspond to the numerical solution of the three-body Schr{\"o}dinger equation with potential (\ref{potential}); so do the large blue dots, using a larger basis set for convergence; the red dashed line is the fit of Eq.~(\ref{E0}) to the data for $L\lambda_0^{1/2}\in[4.5,10]$, including the effective range correction (see text). Inset: same as the main figure, but for $mL^2E/\hbar^2$.}
\label{fig:Energy-Final}
\end{figure}

\begin{figure}[t]
\includegraphics[width=0.5\textwidth]{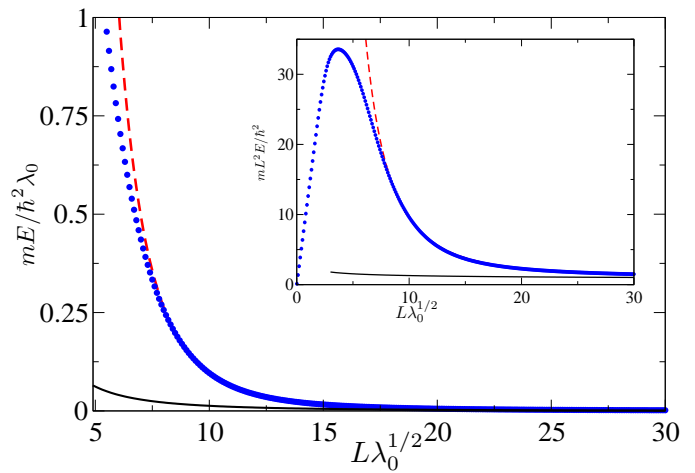}
\caption{Ground state energy of three particles with pairwise interactions $V(x)=W(x)+W(-x)$, with $W(x)$ in Eq.~(\ref{COSH}), with $\lambda_0^{1/2}x_0 = 2.13\ldots$, $\lambda_1/\lambda_0 = 1$, $mV_0/\hbar^2\lambda_0=7$, $V_1/V_0=-0.81047893$, and $b=1$ corresponding to inverse scattering length $x_0/a \approx -6.9\cdot 10^{-7}$. Blue dots correspond to the numerical solution of the three-body Schr{\"o}dinger equation with the full potential; the red dashed line is the fit of Eq.~(\ref{E0}) to the data for $L\lambda_0^{1/2}\in[8,15]$, including the effective range correction (see text); the solid black line corresponds to neglecting the effective range correction. Inset: same as the main figure, but for $mL^2E/\hbar^2$.}
\label{fig:Energy-COSH}
\end{figure}

In the three-body problem under the resonant and no bound state conditions, the contribution of the lowest-order effective three-body force to the ground state energy, Eq.~(\ref{E0}), is dominant. However, higher order effects are present, and in order to extract the three-body momentum scale $Q_*$ accurately, a next-order term of $O(L^{-4})$ must be included. To see what this term corresponds to, I write the three-body effective range correction to the scattering amplitude by simply replacing
\begin{equation}
  \frac{1}{g_R} \to \frac{1}{g_R} - r_3^2k^2,
\end{equation}
which is completely analogous to the problem of 2D two-body scattering \cite{Adhikari2D}. The correction to the energy due to the effective range is given by $\Delta E_{r_3}= 16\pi^4r_3^2g_R^3\hbar^2/mL^4$. This results in a two-parameter fit that needs at least two numerical or experimental data points. In Fig.~\ref{fig:Energy-Final}, I plot the ground state energy of three particles in a periodic box as a function of the system's size for a resonant interaction of the form (\ref{potential}). I extract the ultraviolet (UV) scale $Q_*$ by fitting Eq.~(\ref{E0}), including the next-to-leading order correction $\propto g_R^3L^{-4}$ due to the three-body range, to the numerical data. The agreement between the theory and the data is remarkably good, especially for the rescaled energy $L^2E_0$ (see inset of Fig.~\ref{fig:Energy-Final}, which is a much more stringent test than the energy itself). From Fig.~\ref{fig:Energy-Final}, one sees that the effective three-body interaction is purely repulsive, with fitted values $Q_*x_0=420$ and $r_3/x_0=27.8$. Other choices of the particular functional form of the potential, provided they are repulsive at short distances and form no three-body bound states, and other particular values of the potential's parameters that keep the scattering length divergent yield qualitatively identical results. For instance, in Fig.~\ref{fig:Energy-COSH} I show the results using the potential $V(x)=W(x)+W(-x)$, with $W$ in Eq.~(\ref{COSH}). There, I have chosen parameters that make the potential have a significantly broader and stronger repulsive core at short distances, for which the three-body effective range plays a much larger role at finite energies, having $Q_*x_0=49.3$ and $r_3/x_0=45.5$. In both cases, the ground state energy reaches asymptotically (i.e., at low energy or large $L$) Eq.~(\ref{E0}). 

\section{Experimental considerations}
I now move on to discuss the possible experimental demonstration of the three-body force in one dimension under the two conditions mentioned throughout this work. It must be noted that the requirement of no two-body bound states is given for theoretical convenience. Experimentally, this is justified when there are no shallow bound states, yet with sizeable binding energies, as deep bound states that generically exist in ultracold atomic systems are far in energy from the continuum and are therefore not populated in typical experimental time scales. The resonant condition can be satisfied by either using transversal confinement with anharmonic, anisotropic traps \cite{SaenzCIR}, which can reach effectively infinite scattering lengths \cite{SaenzCIR,ValienteCIR}, magnetic Feshbach resonances \cite{FeshbachReview}, for example for $^{133}\mathrm{Cs}$, which can set the 3D scattering length to zero \cite{Naegerl}, and therefore the effective 1D scattering length to infinity via dimensional reduction \cite{Olshanii}, or a combination thereof.

As for the observation of tangible effects of the three-body forces, I shall consider a realistic experimental scenario, in particular Ref.~\cite{Naegerl}. There, they effectively confine a BEC of $^{133}\mathrm{Cs}$ atoms to one dimension by applying a transversal 2D optical lattice with effective harmonic length $a_{\perp} = 1440 a_0$, with $a_0$ the Bohr radius, and the system consists of an array of quasi-1D tubes. The longitudinal harmonic length is given by $a_{\parallel}=8310a_0$. The central tube, in the repulsive weak coupling regime relevant to this work, has only a few atoms, $N=8$ -- $11$. A 1D resonance ($1/a=0$) is obtained for a magnetic field $B=17.119$ $\mathrm{G}$, at which point they measure the lowest longitudinal breathing mode and show that indeed it corresponds to the non-interacting limit within experimental uncertainty. In the model interaction of Eq.~(\ref{potential}), the most relevant length scale is given by $x_0$, which marks the position of the potential minimum. Typical interatomic interactions have $x_0\sim 5$ -- $10\mathrm{\AA}$ \cite{Aziz}. Using these values for $x_0$, the example given above, which shows a sizeable effect of the three-body force, yet remaining in the weak coupling limit, corresponds to a UV three-body momentum scale $Q_*\sim 42$ -- $84 \mathrm{\AA}^{-1}$ ($Q_*x_0\approx 420$). With the density in the three-body sector $\rho_{3\mathrm{b}}=3/L \sim 10^{-2}$ -- $10^{-1} \mathrm{\AA}^{-1}$, the relevant dimensionless constant $\kappa = Q_*/\rho_{3\mathrm{b}}$ takes on values in the range $\kappa \sim 420$ -- $8400$. Since the three-body force is weak and the particle numbers in the experiment of Ref.~\cite{Naegerl} are small, one can estimate the central density $\rho(0)$ in the central tube by using the non-interacting ground state in a harmonic well. This gives $\rho(0)\approx N/\sqrt{\pi a_{\parallel}^2}$, with values in the range $\rho(0)\sim 3.2 \mu\mathrm{m}^{-1}$ -- $4.4 \mu\mathrm{m}^{-1}$, implying that $Q_*/\rho(0)\sim 10^5$ -- $3 \cdot 10^5$, corresponding to a three-body coupling constant reduced by a factor of about $2$ -- $5$ (see Eq.~(\ref{gR})), very similar in magnitude to the example shown here. It would be, nevertheless, beneficial to have higher particle numbers in the tube, of the order of 100, which would mildly increase the three-body coupling constant. Since the focus of Ref.~\cite{Naegerl} was the strongly interacting limit, it would be interesting to explore the 1D resonant regime in more detail, and study the shift in breathing mode frequency due to the residual three-body interactions. Another possible experimental observation of the effects of three-body forces would be through measurements of the speed of sound, which can be probed using magnetic field gradients \cite{Pan} or Bragg spectroscopy \cite{Hulet} in quasi-1D ultracold atomic systems. Last but not least, an optical lattice may be used to probe three-body interactions, within one well, by means of time-resolved observation of quantum phase revivals \cite{Johnson1,Will}, providing yet another promising platform for experimental verification.

\section{Conclusions}
To conclude, I have studied the three-body problem with identical bosons in one spatial dimension interacting via semi-realistic pairwise interactions and found that, on resonance, the leading order contribution to the three-body scattering amplitude corresponds to an effective, repulsive three-body interaction. I have analyzed the problem in a finite box with periodic boundary conditions, which allows for the extraction of the three-body collisional parameters, and hence the scattering amplitude, via the ground state energy of the system. I have also shown that under rather usual experimental conditions, the effects of the three-body force should be observable. For instance, the leading order correction to the lowest compressional mode in a harmonic trap can be non-negligible even in the weak-coupling regime \cite{ValientePastukhov}. Due to the kinematic equivalence between three-body interactions in 1D and two-body interactions in 2D at low energies, the three-body effective range will change the effects of the anomaly quantitatively, as shown by Hu {\it et al.} in two dimensions, albeit for spin-$1/2$ fermions \cite{HuLianyi}. It is also worth noticing that the resonant condition is not stringent, and these effects are sizeable slightly away from the resonance, even on the slightly attractive side of it. Moreover, the energy shifts due to four- and five-body forces na{\"i}vely scale as $L^{-3}$ and $L^{-4}$, respectively, and may play a non-trivial role in the equation of state of the resonant Bose gas. Many-body physics under these conditions are yet to be explored, and open up a plethora of new possibilities with one-dimensional quantum many-particle systems beyond the Lieb-Liniger model. 

\begin{acknowledgments}
  I would like to thank N.~T. Zinner, R.~E. Barfknecht, M. Mikkelsen and D.~S. Petrov for useful discussions.
\end{acknowledgments}

\appendix

\section{Structure of three-body amplitudes at two-body resonances}\label{apendicitis}
Under the conditions of infinite scattering length and no two-body bound states, the only possible collisional process for three bosons is $3\to 3$ scattering. I show here, using scattering theory, that the residual $3\to 3$ amplitude on resonance at low energies is due to two-body off-shell processes, by showing that the two-body on-shell contributions vanish identically at low energies. I then proceed to show the asymptotic behaviour of the $3\to 3$ scattering states.

\subsection{Faddeev equations}
I briefly review here the derivation of the Faddeev equations (see, e.g. \cite{MarceloDresden}), mainly to establish the notation below unambiguously. The $3\to 3$ transition (T-) matrix $\hat{T}_3(z)$, where $z$ is the (complex) energy, satisfies the Lippmann-Schwinger equation
\begin{equation}
\hat{T}_3(z) = \hat{V}_3+\hat{V}_3\hat{G}_0(z)\hat{T}_3(z),\label{LippmannFaddeev}
\end{equation}
with $\hat{V}_3=\sum_{i=1,2,3}\hat{V}^i$ the interaction potential in the usual notation, and $\hat{V}^i$ the two-body interaction excluding particle $i$ and involving particles $j\ne i$ and $l\ne i$. In Eq.~(\ref{LippmannFaddeev}), $\hat{G}_0(z)$ is the three-body non-interacting Green's function. I rewrite the T-matrix using the Faddeev decomposition
\begin{align}
  \hat{T}_3(z)&=\sum_{i=1,2,3}\hat{T}^i(z)\\
  \hat{T}^i(z)&=\hat{V}^i+\hat{V}^i\hat{G}_0(z)\sum_{s=1,2,3}\hat{T}^s.\label{Faddeev1}
\end{align}
Defining now $\hat{t}^i(z)=(1-\hat{V}^i\hat{G}_0(z))^{-1}\hat{V}^i$, Eq.~(\ref{Faddeev1}) takes the form
\begin{equation}
 \hat{T}^i(z) = \hat{t}^i(z) +\hat{t}^{i}(z)\hat{G}_0(z)\left[\hat{T}^j(z)+\hat{T}^l(z)\right],\label{FaddeevEquations} 
\end{equation}
where $i=1,2,3$ and $i\ne j \ne l \ne i$, which constitutes a set of three coupled integral equations known as Faddeev equations.

The operators $\hat{t}^i(z)$ are off-shell two-body T-matrices \cite{MarceloDresden}. To see this, write their corresponding uncoupled Lippmann-Schwinger equations, which involve only one of the three two-body interactions
\begin{equation}
  \hat{t}^i(z) = \hat{V}^i+\hat{V}^i\hat{G}_0(z)\hat{t}^i(z).\label{LippmannSpectator}
\end{equation}
The difference between the above equation and the usual two-body Lippmann-Schwinger equation is that $\hat{G}_0(z)$ above is the three-body Green's function containing a spectator particle $i$. Denoting $\ket{\mathbf{k}}\equiv \ket{k_1,k_2,k_3}$, since $\bra{\mathbf{k}'}\hat{V}^i\ket{\mathbf{k}}\propto \delta(k_i-k_i')\delta(K-K')$, it is clear that
\begin{equation}
  \bra{\mathbf{k}'}\hat{t}^i(z)\ket{\mathbf{k}}=(2\pi)^2 \delta(k_i-k_i')\delta(K-K')\bra{k_{jl}'}\hat{\tau}_{K,k_i}^i(z)\ket{k_{jl}},
\end{equation}
where $k_{jl}=(k_j-k_l)/2$, $K=k_1+k_2+k_3$, and where I have defined a connected (i.e. free of delta functions) operator $\hat{\tau}_{K,k_i}^i(z)$ which depends parametrically on $K$ and $k_i$. I now define the two-body T-matrix in the relative coordinates at relative energy $\xi$ as $\hat{T}_2(\xi)$. After writing the Lippmann-Schwinger equations for $\hat{T}_2$ and $\hat{\tau}_{K,k_i}$ in the momentum representation, simple comparison yields $\hat{\tau}_{K,k_i}(z) = \hat{T}_2(\xi_i)$, with
\begin{equation}
  \xi_i = z - \frac{3}{4}\frac{\hbar^2 k_i^2}{m} - \frac{\hbar^2K}{m}\left(\frac{K}{4}-k_i\right).\label{xi}
\end{equation}

\subsection{Two-body on-shell contributions to the on-shell three-body amplitude}
The Faddeev equations, Eq.~(\ref{FaddeevEquations}), imply that the three-body T-matrix $\bra{\mathbf{k}'}\hat{T}_3\ket{\mathbf{k}}$ has disconnected contributions ($\propto \sum_{s=1,2,3}\delta(k_s-k_s')\delta(K-K')$) corresponding to two-body scattering, and connected contributions (only $\propto \delta(K-K')$) corresponding to genuine three-body scattering. The disconnected contributions $\hat{T}_3^{\mathrm{nc}}(z)\equiv \sum_{s=1,2,3}\hat{T}^{\mathrm{nc},s}$ vanish on-shell at low energies for infinite scattering length. To see this, notice that
\begin{equation}
  \hat{T}^{\mathrm{nc},i}(z) = \hat{t}^{i}(z).
\end{equation}
In the previous subsection, I showed that $\bra{\mathbf{k}'}\hat{t}^{i}(z)\ket{\mathbf{k}} = (2\pi)^2\delta(K-K')\delta(k_i-k_i')\hat{T}_2(\xi_i)$, with $\xi_i$ given in Eq.~(\ref{xi}). The disconnected part of the reduced component $i$ of the T-matrix (i.e. after dropping the delta function ensuring total momentum conservation) takes on the value
\begin{equation}
  \langle k_{ij}',k_{jl}'||\hat{T}_K^{\mathrm{nc},i}(z)|| k_{ij}k_{jl}\rangle = 2\pi \delta(k_i-k_i')\bra{k_{jl}'}\hat{T}_2(\xi_i)\ket{k_{jl}}.\label{Disconnected}
\end{equation}
Because of Galilean invariance, I shall set $K=0$ without loss of generality from now on, and drop its dependence. To go on the energy shell, I set
\begin{align}
  z &=\frac{4}{3}\frac{\hbar^2}{m}\left(k_{ij}^2+k_{jl}^2+k_{ij}k_{jl}\right)+i\eta\nonumber\\
&=
\frac{4}{3}\frac{\hbar^2}{m}\left(k_{ij}'^2+k_{jl}'^2+k_{ij}'k_{jl}'\right)+i\eta\label{zOnShell}
\end{align}
with $\eta\to 0^+$. The low-energy limit ($z\to 0$) implies,therefore, both simultaneous limits $k_{ij}\to 0$ and $k_{jl}\to 0$. The on-shell two-body T-matrix at low energies admits the following effective range expansion \cite{Adhikari1D}
\begin{equation}
  \bra{\pm k_{jl}}\hat{T}_2(\hbar^2k_{jl}^2/m+i\eta)\ket{k_{jl}} \approx \frac{2\hbar^2/m}{-\frac{a}{1+\frac{1}{2}r_2ak_{jl}^2}+i/|k_{jl}|},\label{OnShell}
\end{equation}
where $a$ and $r_2$ are, respectively, the two-body scattering length and effective range. Note that, using Eqs.~(\ref{xi}) and (\ref{zOnShell}),  $\xi_i = \hbar^2k_{jl}^2/m+i\eta$. Therefore, the disconnected part of the Faddeev component $i$, Eq.~(\ref{Disconnected}), corresponds to the on-shell two-body T-matrix, Eq.~(\ref{OnShell}). On resonance ($1/a = 0$), Eq.~(\ref{OnShell}) becomes
\begin{equation}
  \bra{\pm k_{jl}}\hat{T}_2(\hbar^2k_{jl}^2/m+i\eta)\ket{k_{jl}} \approx \frac{2\hbar^2k_{jl}^2/m}{-2/r_2+i|k_{jl}|} = O(z).
\end{equation}
Because of reduced dimensionality, it is not sufficient to show that the disconnected contribution vanishes for $z\to 0$, since at low energies the inverse two-body T-matrix is always infrared divergent, but I also need to show that it vanishes faster than for finite scattering length. This is easy to see from Eq.~(\ref{OnShell}), which, for non-zero $1/a$ behaves as $\propto |k_{jl}|$ , i.e. the on-shell two-body T-matrix is of $O(z^{1/2})$, as I wanted to prove.

It only remains now to be shown that the two-body on-shell contributions to the connected part of the three-body T-matrix also vanish on-shell at low energies. To see this, introduce the disconnected part of the (off-shell) components $\hat{T}^{\mathrm{nc},j}$ and $\hat{T}^{\mathrm{nc},l}$ into the Faddeev equation for the component $i$, Eq.~(\ref{FaddeevEquations}). For instance, component $j$ gives the following contribution $\Delta_{j}^{\mathrm{c},i}(z)$ to the connected part of component $i$ 
\begin{align}
  &\bra{k_{ij}'k_{jl}'}\Delta_{j}^{\mathrm{c},i}\ket{k_{ij}k_{jl}} \nonumber\\
  &= \bra{k_{jl}'}\hat{T}_2(\xi_i)\ket{\frac{2}{3}(k_{jl}-k_{ij})+\frac{1}{3}(k_{jl}'+2k_{ij}')}\nonumber\\
  &\times  \bra{\frac{2}{3}(k_{jl}'+2k_{ij}')+\frac{1}{3}(k_{jl}+2k_{ij})}\hat{T}_2(\xi_j)\ket{k_{ij}+k_{jl}}\nonumber \\
  &\times \frac{1}{z-\frac{4\hbar^2}{9m}\mathcal{F}}, \label{UglyStuff}
\end{align}
with
\begin{equation}
  \mathcal{F}\equiv (k_{jl}'+2k_{ij}')^2+(k_{jl}-k_{ij})^2+(k_{jl}'+2k_{ij}')(k_{jl}-k_{ij})
\end{equation}
A similarly ugly, yet functionally identical expression holds for the contribution from component $l$. On the energy shell, the denominator in Eq.~(\ref{UglyStuff}) is of $O(z)$. The numerator, being the product of two two-body T-matrices, is of $O(z^2)$. Therefore, on shell, $\bra{k_{ij}'k_{jl}'}\Delta_{j}^{\mathrm{c},i}\ket{k_{ij}k_{jl}}=O(z)$, as I wanted to show. Note that, for non-vanishing $1/a$, the numerator would instead be of $O(z)$, and therefore the contribution $\Delta_{j}^{\mathrm{c},i}$ from on-shell two-body scattering would be finite.

\subsection{Three-body coupling constant}\label{SectionCoupling}
From the above analysis, it is now obvious that any contribution to the on-shell $3\to 3$ amplitude at low energy stems from off-shell two-body processes, over which one integrates in the Faddeev integral equations. At low energy, moreover, I have shown that the only non-vanishing contributions (i.e. of order lower than $O(z)$), if there are any, must be connected. Therefore, it is possible to choose now an arbitrary (but low energy) scale $\mu$, with $z=\mu+i\eta$, and solve the Faddeev equations for the $3\to 3$ amplitude at that energy. At low energies, retaining only terms of lower order than $O(z)$, the amplitude is approximated by a constant, independent of angular variables, by
\begin{equation}
\bra{\mathbf{k}'}\hat{T}_3(\mu+i\eta)\ket{\mathbf{k}}\approx \delta(K-K')T_3^{\mathrm{c},\mathrm{on}}(\mu)+O(z).\label{RenormalisationCondition}
\end{equation}
Defining $1/\tilde{g}_3(\mu)=\mathrm{Re}[1/T_3^{\mathrm{c},\mathrm{on}}(\mu)]$ (its imaginary part is fixed by unitarity), and dropping terms of $O(z)$ and higher, Eq.~(\ref{RenormalisationCondition}) represents the renormalization condition for the lowest-order effective theory of the three-body problem at infinite two-body scattering length.

\subsection{Asymptotics of three-body scattering states}
The above analysis showed that, to $O(z)$, the asymptotic part of the three-body scattering states associated with the disconnected part of the three-body T-matrix corresponds to a non-interacting three-body state. Here, I show that the rest of the scattered wave can be asymptotically described by a pure three-body force.

I begin with the Lippmann-Schwinger equation for the scattering wave function $\ket{\psi}$
\begin{equation}
  \ket{\psi} = \ket{\mathbf{k}} + \hat{G}_0(z)\hat{V}\ket{\psi},
\end{equation}
where $z=E+i\eta$ and $E=\hbar^2\mathbf{k}^2/2m$ (on-shell condition). I assume $\ket{\mathbf{k}}$ has been symmetrised to represent bosons. I employ the Faddeev decomposition for the scattering state, $\ket{\psi}=\ket{\psi^i}+\ket{\psi^j}+\ket{\psi^l}$, with obvious notation. Defining the spectator Green's functions $\hat{G}_i(z)$ as
\begin{equation}
  \hat{G}_i(z) = \hat{G}_0(z)+\hat{G}_0(z)\hat{V}^i\hat{G}_i(z),
\end{equation}
and the disconnected scattering states $\ket{\psi^i_0}$ as
\begin{equation}
  \ket{\psi^i_0} = \hat{S}_3\left[\ket{k_i}(1+\hat{G}_0(z)\hat{t}^i(z))\ket{k_jk_l}\right],\label{LippmannGreen}
\end{equation}
with $\hat{S}_3$ the symmetrization operator for three particles, the Faddeev equations read
\begin{equation}
  \ket{\psi^i} = \ket{\psi^i_0}  + \hat{G}_i(z)\hat{V}^i\left[\ket{\psi^j}+\ket{\psi}^l\right].\label{FaddeevPsi}
\end{equation}
For going into the position representation, I use Jacobi coordinates $X=(x_1+x_2+x_3)/3$, $x=(x_1-x_2)/\sqrt{2}$ and $y=\sqrt{2/3}[x_3-(x_1+x_2)/2]$. The scattering states have well-defined total momentum $K$, and have the form
\begin{equation}
  \psi^i(X,x,y) = e^{iKX}\psi^i_K(x,y).
\end{equation}
In the following, I will drop the subscript $K$ from $\psi^i_K$ when there is no room for confusion, and set $K=0$ without loss of generality. The non-interacting three-body Green's function, after separation of the centre of mass coordinate, satisfies
\begin{equation}
  z+\frac{\hbar^2}{2m}(\partial^2_x+\partial^2_y)G_0(z;\mathbf{r}_{\perp},\mathbf{r}'_{\perp})=\delta^{(2)}(\mathbf{r}_{\perp}-\mathbf{r}'_{\perp}),\label{GreenEquation}
\end{equation}
where $z$ should be replaced by $z-\hbar^2K^2/6m$ for non-zero momentum, $\mathbf{r}_{\perp}=(x,y)$, and where I have defined $\bra{\mathbf{r}_{\perp}}\hat{G}_0(z)\ket{\mathbf{r}'_{\perp}}\equiv G_0(z;\mathbf{r}_{\perp},\mathbf{r}'_{\perp})$. From Eq.~(\ref{GreenEquation}), it is obvious that $\hat{G}_0(z)$ is the usual two-dimensional, single-particle, non-interacting, non-relativistic Green's function \cite{Adhikari2D}, given by
\begin{equation}
  G_0(z;\mathbf{r}_{\perp},\mathbf{r}'_{\perp}) = -\frac{2m}{\hbar^2}\frac{i}{4}H_0^{(1)}(k|\mathbf{r}_{\perp}-\mathbf{r}'_{\perp}|),
\end{equation}
where $k^2=k_x^2+k_y^2$ ($k>0$), and $H_0^{(1)}$ is the zero-th order Hankel function of the first kind. Asymptotically ($|\mathbf{r}_{\perp}|\equiv r_{\perp}\to \infty$), the Green's function takes the form \cite{Adhikari2D}
\begin{equation}
  G_0(z;\mathbf{r}_{\perp},\mathbf{r}'_{\perp}) \to -\frac{2m}{\hbar^2}\frac{e^{i\pi/4}}{4}\sqrt{\frac{2}{\pi kr_{\perp}}}e^{ikr_{\perp}}e^{-i\mathbf{k}\cdot \mathbf{r}'_{\perp}}.\label{G0Asymptote}
\end{equation}
Introducing Eq.~(\ref{G0Asymptote}) into the position representation of the Lippmann-Schwinger equation for the spectator Green's function, Eq.~(\ref{LippmannGreen}), I find, for $r_{\perp}\to \infty$,
\begin{equation}
  G_i(z;\mathbf{r}_{\perp},\mathbf{r}'_{\perp})\to G_0(z;\mathbf{r}_{\perp},0)\chi_i(z;\mathbf{r}'_{\perp}),\label{AsymptoticGi}
\end{equation}
with
\begin{equation}
  \chi_i(z;\mathbf{r}'_{\perp})= 1 + \int d\mathbf{r}''_{\perp} e^{-i\mathbf{k}\cdot\mathbf{r}''_{\perp}}V^i(\mathbf{r}''_{\perp})G_i(z;\mathbf{r}''_{\perp},\mathbf{r}'_{\perp}).
\end{equation}
Separating the Faddeev components as $\ket{\psi^i}=\ket{\psi^i_0}+\ket{\phi^i}$, with $\ket{\phi^i}$ the connected part of the scattered wave, and inserting the asymptotic form (\ref{AsymptoticGi}) of the Green's function into the Faddeev equations (\ref{FaddeevPsi}), I obtain
\begin{equation}
  \phi^i(\mathbf{r}_{\perp})\to \beta_i(z;\theta,\theta_r)G_0(z;\mathbf{r}_{\perp},0), \hspace{0.2cm} r_{\perp}\to \infty,
\end{equation}
with
\begin{equation}
  \beta_i(z,\theta,\theta_r) = \int d\mathbf{r}' \chi_i(z;\mathbf{r}'_{\perp})V^i(\mathbf{r}'_{\perp})\left[\psi^j(\mathbf{r}'_{\perp})+\psi^l(\mathbf{r}'_{\perp})\right].\label{beta}
\end{equation}
Note that, above, the dependence on the angles $\theta$ and $\theta_r$ ($k_x=k\cos\theta$, $k_y=k\sin\theta$, $x=r_{\perp}\cos\theta_r$, $y=r_{\perp}\sin\theta_r$) has been made explicit and depends on the incident state $\ket{\mathbf{k}}$ via the Faddeev components $\psi^i$ entering the integral in Eq.~(\ref{beta}). As is well-known in scattering theory \cite{Taylor}, the quantity $\beta_i(z,\theta,\theta_r)$ is proportional to the connected part of the on-shell T-matrix, $\beta_i(z;\theta,\theta_r)\propto \bra{k,\theta_r}T_3^{\mathrm{conn.}}(z)\ket{k,\theta}$.

This long detour shows that, if $\beta_i$, and therefore the connected part of the three-body T-matrix, vanishes slower than $O(z)$ as $z\to 0$, then any significant low-energy scattering is due to genuine three-body processes involving off-shell two-body amplitudes, i.e. these are dominant at low energies. In particular, if a three-boson system is placed on a finite line with periodic boundary conditions, as was done above, any non-zero shift in the ground state energy is solely due to these processes. Last but not least, unitarity requires, since the connected part of the three-body problem is two dimensional at long distances \cite{Adhikari2D}, that if this part of the T-matrix is dominant, at low energies (for which scattering becomes isotropic), then it vanishes logarithmically as $\propto 1/\log(E/E_0)$, with $E_0$ an arbitrary energy scale. This has an immediate consecuence regarding the running of the coupling constant in Subsect. \ref{SectionCoupling}. For $z=\mu+i\eta$, using Eq.~(\ref{RenormalisationCondition}), and the unitarity condition, it is not difficult to obtain the coupling constant $\tilde{g}_3(E)$ at any low energy, given $\tilde{g}_3(\mu)$, and the energy scale $E_0$, as
\begin{equation}
\tilde{g}_3(E)=\frac{\tilde{g}_3(\mu)}{1+\frac{\log(E/\mu)}{\log(\mu/E_0)}}.
\end{equation}
The running of the coupling constant above is achieved via an effective three-body interaction as discussed earlier.

\bibliographystyle{unsrt}

\begin{thebibliography}{99}
\bibitem{Yukawa} H. Yukawa, Proc. Phys. Math. Soc. Japan {\bf 17}, 48 (1935).

  
\bibitem{Erkelenz} K. Erkelenz, Phys. Rep. {\bf 13C}, 191 (1974).

\bibitem{Bryan} R.~A. Bryan and B.~L. Scott, Phys. Rev. {\bf 135}, B434 (1964).

\bibitem{Bomelburg} A. B{\"o}melburg, Phys. Rev. C {\bf 28}, 403 (1983).
  
\bibitem{ReviewThreeBody} H.~-W. Hammer, A. Nogga and A. Schwenk, Rev. Mod. Phys. {\bf 85}, 197 (2013).

\bibitem{Nogga} A. Nogga, D. H{\"u}ber, H. Kamada and W. Gl{\"o}cke, Phys. Lett. B {\bf 409}, 19 (1997).

\bibitem{Epelbaum1} E. Epelbaum, H.~-W. Hammer and U.~-G. Mei{\ss}ner, Rev. Mod. Phys. {\bf 81}, 1773 (2009). 

\bibitem{Epelbaum2} E. Epelbaum, H. Krebs and U.~G. Mei{\ss}ner, Phys. Rev. Lett. {\bf 115}, 122301 (2015).
  
\bibitem{WeinbergEFT1} S. Weinberg, Phys. Lett. B {\bf 251}, 288 (1990).

\bibitem{WeinbergEFT2} S. Weinberg, Nucl. Phys. B {\bf 631}, 447 (1998).

\bibitem{Kievsky} A. Kievsky, A. Polls, B. Juli{\'a}-D{\'i}az and N.~K. Timofeyuk, Phys. Rev. A {\bf 96}, 040501(R) (2017).
  
\bibitem{BraatenEFT} E. Braaten, M. Kusunoki and D. Zhang, Ann. Phys. (NY) {\bf 323}, 1770 (2008).

\bibitem{HuangYang} K. Huang and C.~N. Yang, Phys. Rev. A {\bf 105}, 767 (1957).

\bibitem{BEC1} M.~H. Anderson, J.~R. Ensher, M.~R. Matthews, C.~E. Wieman and E.~A. Cornell, Science {\bf 269}, 198 (1995).

\bibitem{BEC2} K.~B. Davis, M.~-O. Mewes, M.~R. Andrews, N.~J. van Druten, D.~S. Durfree, D.~M. Kurn and W. Ketterle, Phys. Rev. Lett. {\bf 75}, 3969 (1995).

\bibitem{Bedaque1} P.~F. Bedaque, H.~-W. Hammer and U. van Kolck, Phys. Rev. Lett. {\bf 82}, 463 (1999).

\bibitem{Bedaque2} P.~F. Bedaque, H.~-W. Hammer and U. van Kolck, Nucl. Phys. A {\bf 646}, 444 (1999).

\bibitem{Grimm} T. Kraemer {\it et al.}, Nature {\bf 440}, 315 (2006).

\bibitem{Grimm2} S. Knoop {\it et al.}, Nature Phys. {\bf 5}, 227 (2009).

\bibitem{Zaccanti} M. Zaccanti {\it et al.}, Nature Phys. {\bf 5}, 586 (2009).

\bibitem{Efimov} V. Efimov, Phys. Lett. B {\bf 33}, 563 (1970).

\bibitem{Bulgac} A. Bulgac, Phys. Rev. Lett. {\bf 89}, 050402 (2002).

\bibitem{Tarruell} C.~R. Cabrera, L. Tanzi, J. Sanz, B. Naylor, P. Thomas, P. Cheiney and L. Tarruell, Science {\bf 359}, 301 (2018).

\bibitem{Inguscio} G. Semeghini {\it et al.}, Phys. Rev. Lett. {\bf 120}, 235301 (2018).

\bibitem{Yin} X.~Y. Yin, D. Blume, P.~R. Johnson and E. Tiesinga, Phys. Rev. A {\bf 90}, 043631 (2014).

\bibitem{Johnson1} P.~R. Johnson, E. Tiesinga, J.~V. Porto and C.~J. Williams, New J. Phys. {\bf 11}, 093022 (2009).

\bibitem{Johnson2} P.~R. Johnson, D. Blume, X.~Y. Yin, W.~F. Flynn and E. Tiesinga, New J. Phys. {\bf 14}, 053037 (2012); {\it ibid.}, New J. Phys. {\bf 20}, 079501 (2018). 

\bibitem{Will} S. Will, T. Best, U. Schneider, L. Hackerm{\"u}ller, D.~-S. L{\"u}mann and I. Bloch, Nature {\bf 465}, 09036 (2010).

\bibitem{Dobr} J. Dobrzyniecki, X. Li, A.~E.~B. Nielsen and T. Sowi{\'n}ski, Phys. Rev. A {\bf 97}, 013609 (2018).

\bibitem{Bermudez} A. Bermudez, D. Porras, M.~A. Martin-Delgado, Phys. Rev. A {\bf 79}, 060303 (2009).
  
\bibitem{PetrovThreeBody} G. Guijarro, A. Pricoupenko, G.~E. Astrakharchik, J. Boronat and D.~S. Petrov, Phys. Rev. A {\bf 97}, 061605 (2018).

\bibitem{Drut1} J.~E. Drut, J.~R. McKenney, W.~S. Daza, C.~L. Lin and C.~R. Ord{\'o}{\~{n}}ez, Phys. Rev. Lett. {\bf 120}, 243002 (2018).

\bibitem{Drut2} W.~S. Daza, J.~E. Drut, C.~L. Lin and C.~R. Ord{\'o}{\~{n}}ez, e-print arXiv:1808.0711v1 .

\bibitem{Nishida2} Y. Nishida, Phys. Rev. A {\bf 97}, 061603 (2018).
  
\bibitem{Nishida3BodyLattice} Y. Sekino and Y. Nishida, Phys. Rev. A {\bf 97}, 011602 (2018).

\bibitem{Pricoupenko1} L. Pricoupenko, Phys. Rev. A {\bf 97}, 061604 (2018).
  
\bibitem{Pricoupenko} L. Pricoupenko, Phys. Rev. A {\bf 99} , 012711 (2019).

\bibitem{PastukhovThreeBody} V. Pastukhov, Phys. Lett. A (in press), https://doi.org/10.1016/j.physleta.2018.12.006 (2018).

\bibitem{Harshman} N.~L. Harshman and A.~C. Knapp, e-print arXiv:1803.11000v2 .

\bibitem{anyons} F. Wilczek, Phys. Rev. Lett. {\bf 49}, 957 (1982).
  
\bibitem{Selim} M. Holten, L. Bayha, A.~C. Klein, P.~A. Murthy, P.~M. Preiss and S. Jochim, Phys. Rev. Lett. {\bf 121}, 120401 (2018).

\bibitem{Vale} T. Peppler, P. Dyke, M. Zamorano, S. Hoinka and C.~J. Vale, Phys. Rev. Lett. {\bf 121}, 120402 (2018).  

\bibitem{AdrianDelMaestro1} A. Del Maestro, M. Boninsegni and I. Affleck, Phys. Rev. Lett. {\bf 106}, 105303 (2011).

\bibitem{AdrianDelMaestro2} A. Del Maestro, Int. J. Mod. Phys. B {\bf 26}, 1244002 (2012).

\bibitem{AdrianDelMaestro3} P.~F. Duc, M. Savard, M. Petrescu, B. Rosenow, A. Del Maestro and G. Gervais, Science Adv. {\bf 1}, e1400222 (2015).
  
\bibitem{LiebLiniger} E.~H. Lieb and W. Liniger, Phys. Rev. {\bf 130}, 1605 (1963).
  
\bibitem{ValienteZinnerEFT} M. Valiente and N.~T. Zinner, Few-body syst. {\bf 56}, 845 (2015).

\bibitem{Luscher1} M. L{\"u}scher, Commun. Math. Phys. {\bf 104}, 177 (1986).

\bibitem{Luscher2} M. L{\"u}scher, Commun. Math. Phys. {\bf 105}, 153 (1986).

\bibitem{BeaneTwoNucleons} S.~R. Beane, Phys. Lett. B {\bf 585}, 106 (2004).

\bibitem{LatticeQCDReview} R.~A. Brice{\~n}o, J.~J. Dudek and R.~D. Young, Rev. Mod. Phys. {\bf 90}, 025001 (2018). 
  
\bibitem{Adhikari1D} V.~E. Barlette, M.~M. Leite and S. Adhikari, Eur. J. Phys. {\bf 21}, 435 (2000).

  
\bibitem{Beane2D} S.~R. Beane, Phys. Rev. A {\bf 82}, 063610 (2010).

\bibitem{ValienteZinnerL} M. Valiente and N.~T. Zinner, Sci. China-Phys. Mech. \& Astr. {\bf 59}, 114211 (2016).
  
\bibitem{Adhikari2D} S. Adhikari, Am. J. Phys. {\bf 54}, 362 (1986).

\bibitem{SaenzCIR} S. Sala, P.~I. Schneider and A.~S. Saenz, Phys. Rev. Lett. {\bf 109}, 073201 (2012).

\bibitem{ValienteCIR} M. Valiente and K. M{\o}lmer, Phys. Rev. A {\bf 84}, 053628 (2011). 

  
\bibitem{FeshbachReview} C. Chin, R. Grimm, P. Julienne and E. Tiesinga, Rev. Mod. Phys. {\bf 82}, 1225 (2010).

\bibitem{Naegerl} E. Haller {\it et al.}, Science {\bf 325}, 1224 (2009).

\bibitem{Olshanii} M. Olshanii, Phys. Rev. Lett. {\bf 81}, 938 (1998).

\bibitem{Aziz} R.~A. Aziz, V.~P.~S. Nain, J.~S. Carley, W.~L. Taylor and
G.~T. McConville, J. Chem. Phys. {\bf 70}, 4330 (1979).
  
\bibitem{Pan} B. Yang {\it et al.}, Phys. Rev. Lett. {\bf 119}, 165701 (2017).

\bibitem{Hulet} T.~L. Yang {\it et al.}, Phys. Rev. Lett. {\bf 121}, 103001 (2018).

\bibitem{ValientePastukhov} M. Valiente and V. Pastukhov, Phys. Rev. A {\bf 99}, 053607 (2019).

\bibitem{HuLianyi} H. Hu, B.~C. Mulkerin, U. Toniolo, L. He and X.-~J. Liu, Phys. Rev. Lett. {\bf 122}, 070401 (2019).

\bibitem{MarceloDresden} M.~T. Yamashita, D.~S. Rosa and J.~H. Sandoval, Few-body syst. {\bf 59}, 19 (2018).  

\bibitem{Taylor} J.~R. Taylor, {\it Scattering Theory. The Quantum Theory of Nonrelativistic Collisions} (Dover publications, New York 2006).
  
\end{thebibliography}

\end{document}